\documentclass[apj]{emulateapj}
\usepackage{color}
\usepackage{graphicx} 
\usepackage{subfigure} 
\usepackage{amsmath} 
\usepackage{hyperref} 
\usepackage[utf8]{inputenc} 

\def\tninety{\ensuremath{T_{90}}}
\def\fluence{\ensuremath{S_{\gamma}}}
\def\fx{\ensuremath{F_\textsc{x}^{11h}}}
\def\fr{\ensuremath{F_\textsc{r}^{11h}}}
\def\ebol{\ensuremath{E_{iso}^{bol}}}
\def\eke{\ensuremath{E_{\textsc{k},iso}}}
\def\epsgamma{\ensuremath{\epsilon_\gamma}}

\def\miriad{{\sc miriad}}
\def\aips{{\sc aips}}

\def\mJy{\ensuremath{\mathrm{mJy}}}
\def\swift{{\em Swift}}
\renewcommand\bf{{}}

\begin{document}
\title{Two populations of gamma-ray burst radio afterglows}

\author{P. J. Hancock\altaffilmark{1}\thanks{E-mail:Paul.Hancock@Sydney.edu.au}, B. M. Gaensler\altaffilmark{1} and T. Murphy\altaffilmark{1,2}}
\affil{Sydney Institute for Astronomy (SIfA), School of Physics, The University of Sydney, NSW 2006, Australia}
\altaffiltext{1}{ARC Centre of Excellence for All-sky Astrophysics (CAASTRO)}
\altaffiltext{2}{School of Information Technologies, The University of Sydney, NSW 2006, Australia}

\begin{abstract}
The detection rate of gamma-ray burst (GRB) afterglows is only
$\sim30\%$ at radio wavelengths, much lower than in the X-ray
($\sim95\%$) or optical ($\sim70\%$) bands. The cause of this low
radio detection rate has previously been attributed to limited
observing sensitivity. We use visibility stacking to
test this idea, and conclude that the low detection rate is instead
due to two intrinsically different populations of GRBs, radio bright
and radio faint. {\bf We calculate that no more than 70\% of GRB afterglows are
truly radio bright, leaving a significant population of GRBs that lack a radio
afterglow.} These radio bright GRBs have higher gamma-ray
fluence, isotropic energies, X-ray fluxes and optical fluxes than the
radio faint GRBs, confirming the existence of two physically distinct
populations. We suggest that the gamma-ray efficiency of the prompt
emission is responsible for the difference between the two
populations. We also discuss the implications for future radio and
optical surveys.
\end{abstract}


\section{Introduction}\label{sec:intro}
The standard gamma-ray burst (GRB) afterglow model
\citep{piran_gamma-ray_1999,woosley_supernova-gamma-ray_2006} describes the
afterglow as an expanding fireball. The shape and evolution of the
afterglow spectrum contain a number of spectral and temporal breaks
that depend on the environment into which the ejecta are expanding and
on the micro-physical properties of the shock. The radio afterglow is
a product of the GRB ejecta interacting with the circumstellar
material.

The \swift{} satellite \citep{gehrels_swift_2004} was the first
mission that could provide fast localization of GRBs good enough that
ground based optical follow up could be obtained for a large number of
bursts. However even after many years of ground based optical and IR
follow up, only $\sim50\%$ of GRBs had a detectable optical afterglow,
with the optically non-detected GRBs labeled as ``dark'' GRBs
\citep{jakobsson_swift_2004}. The difference between the dark and
normal GRBs was eventually found to be a combination of extrinsic
factors (extinction, redshift, and observing delay) rather than
intrinsic factors such as luminosity \citep{greiner_nature_2011}. When
observations are begun within 4 hours of the burst, optical afterglows
are detected $90\%$ of the time \citep{greiner_nature_2011}.

At radio wavelengths the detection rate of GRB afterglows is even
lower \citep[$\sim 30\%$,][]{chandra_radio-selected_2012} than at optical or
X-ray wavelengths. It has been generally accepted that the low
detection rates are due to instrumental sensitivity
\citep[eg,][]{frail_radio_2005}, however this cause has not yet been
tested experimentally.

\section{GRB radio afterglows}
\begin{figure}
\centering
\includegraphics[width=\linewidth]{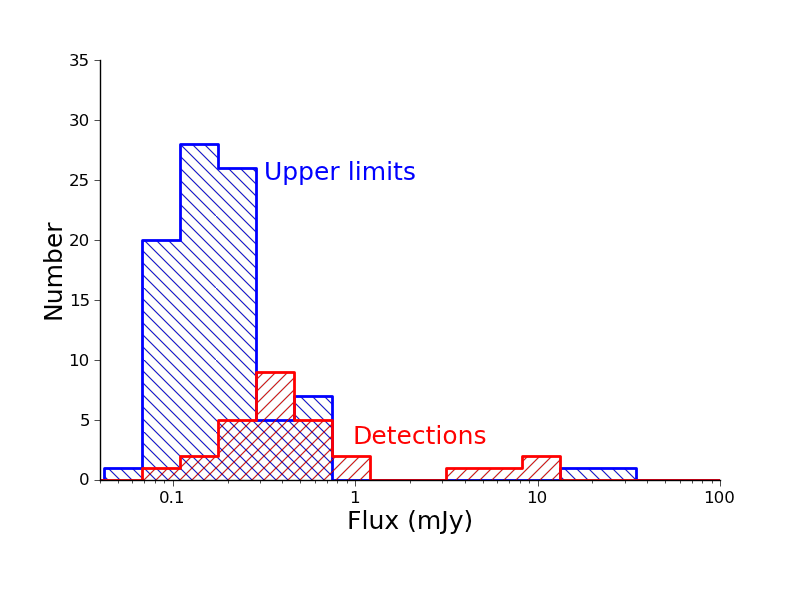}
\caption{The measured flux or $3\sigma$ upper limit for a sample of
GRBs observed with the VLA within the first 5 days after the burst at
a frequency of $8.46$\,GHz. The $3\sigma$ upper limits and detections
are not very well separated, consistent with the claim made by
\citet{chandra_radio-selected_2012} -- that the detection rate of GRB
afterglows is limited by instrumental sensitivity.}
\label{fig:detectionrate}
\end{figure}

In a recent review of the radio properties of GRB afterglows,
\citet{chandra_radio-selected_2012} present a large archival sample of
radio observations of GRBs. Despite the large number of radio
observations (2995), only 95 of the 304 GRBs observed had a confirmed
radio afterglow. In their review, \citet{chandra_radio-selected_2012}
point out that the upper limits on and detected fluxes of radio
afterglows are not significantly
different. Figure\,\ref{fig:detectionrate} shows the distribution of
detected fluxes and $3\sigma$ upper limits for GRB radio afterglows at
$8.46$\,GHz in the first five days after the burst.

We refer to GRBs with detected radio afterglows as {\em radio bright}
GRBs, and those without detected radio afterglows as {\em radio faint}
GRBs. Once can consider three possible explanations for the low
detection rate of GRB radio afterglows: redshift, observing
sensitivity, or intrinsic differences between two sub-types of GRB. If
we assume that the radio bright and radio faint GRB samples are
intrinsically the same but at different redshifts then we would expect
that the bright GRBs are bright only because they are nearer to us
than the faint GRBs. It is therefore possible that the difference
between the bright and faint samples is simply an artifact of their
different redshift distributions. If the redshift distributions of the two
populations are the same, then a population of GRBs with an
intrinsically broad luminosity distribution would be artificially
divided into two populations of radio bright and radio faint GRBs
simply because of limited observing sensitivity. In this situation the
low detection rate is nothing more than an artifact of limited
sensitivity.

However, if the detection of GRB radio afterglows is biased by
observing sensitivity then it should be possible to extract the mean
afterglow flux using visibility stacking
\citep{hancock_visibility_2011}. In this paper we perform visibility
stacking on observations from the Very Large Array (VLA) in order to
determine the extent to which observing sensitivity is responsible for
observed differences between the radio bright and radio faint
samples. In \S\ref{sec:redshifts} we show that the redshift
distributions of the radio bright and radio faint samples of GRB
afterglows are the same. In \S\ref{sec:sensitivity} we test whether
observing sensitivity can explain the difference between the two
samples; we find that it cannot and that the two samples of GRBs
represent physically distinct populations. In
\S\ref{sec:multi-wavelength} we show that the two populations also have
distinct properties at other wavelengths, and in
\S\ref{sec:interpretation} we suggest a possible cause of the
intrinsic differences between the radio bright and radio faint GRBs.

\section{The redshift distributions of GRBs}\label{sec:redshifts}
\begin{figure}
  \centering
\includegraphics[width=0.95\linewidth]{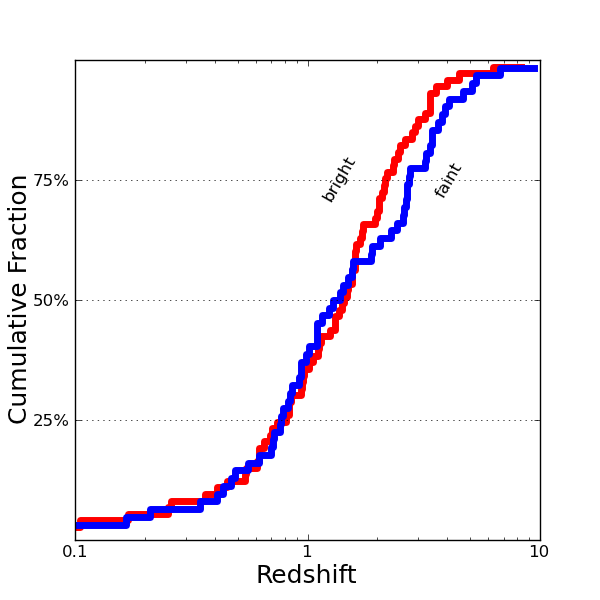}
\caption{The cumulative distribution function of redshifts for both
radio bright (in red) and radio faint (in blue) GRBs. There is no
significant difference between the two populations.}
\label{fig:cdf-z-dnd}
\end{figure}

\citet{chandra_radio-selected_2012} describe a sample of 2995 flux
density measurements and upper limits for 304 GRBs between $0.6$ and
$660$\,GHz with the majority at $8.46$\,GHz. The observations were
taken with the VLA and the Australia Telescope
Compact Array (ATCA) and were of GRBs with a burst date between 1997
and 2011. Table\,1 of that paper lists: redshift, duration ($T_{90}$),
the gamma-ray fluence (\fluence), X-ray flux scaled to 11h post burst
(\fx), and the optical flux scaled to 11h post burst (\fr), for each
of the GRBs that were observed. We used the redshifts listed in this
table to construct a cumulative distribution function for the radio
bright and radio faint GRB samples. The cumulative distribution is
shown in Figure\,\ref{fig:cdf-z-dnd}. A two population K-S test
confirms that the two distributions are not significantly different
($p=0.32$). Thus for GRBs with a known redshift, the radio bright and
radio faint samples have the same distribution of redshifts.

\begin{figure}
\centering
\includegraphics[width=0.95\linewidth]{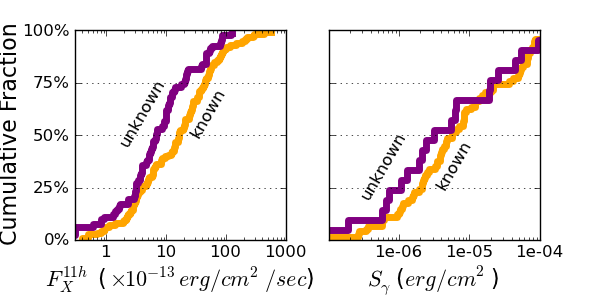}
\caption{Cumulative distribution functions for GRBs with known
(orange) and unknown (purple) redshifts. Shown here are the most and
least different distributions drawn from a wider sample of considered
parameters. {\em Left}: \fx\, for the 178 GRBs with radio observations
($p=0.05$). {\em Right}: \fluence\, for the 70 GRBs in the radio
bright sample ($p=0.88$).}
\label{fig:cdf-extreme}
\end{figure}

The fraction of radio bright GRBs with a known redshift (72\%) is
different from that of radio faint GRBs (45\%). This difference in
knowledge of redshifts could potentially cause biases in the
distribution of other observed properties of GRBs. In order to
evaluate the significance of any such bias we computed cumulative
distribution functions for gamma-ray, X-ray, and optical properties of
the GRBs that do and do not have a
measured redshift. We again make use of Table\,1 of
\citet{chandra_radio-selected_2012} to obtain \fluence{}, \fx{}, and
\fr{}. For each of these parameters a two population K-S test was
carried out between the GRBs with and without known redshifts. These
tests were performed on the full GRB sample, and also on the radio
bright and radio faint sub-samples. In Figure\,\ref{fig:cdf-extreme}
we show the most and least significant differences between the
aforementioned parameter distributions. The resulting p-statistics
from the K-S tests correspond to differences with a significance $\leq
3\sigma$, indicating that the distribution of the aforementioned
properties are not being biased by the presence or lack of a measure
redshift. Each of the aforementioned parameters (even \fluence) are
not available for all GRBs due to selection effects that are beyond
the scope of this work.

The comparison presented in this section shows that the radio bright
and radio faint GRBs have the same redshift distribution. Thus we can
rule out redshift as a cause of the observed difference in radio
brightness. Similarly, our incomplete knowledge of redshift is not
introducing differences in \fluence, \fx, or \fr\, between the two
samples.

\section{The radio flux distribution of GRBs}\label{sec:sensitivity}
In this section we explore the possibility that the distinction
between radio bright and radio faint GRB afterglows is an artifact of
observational sensitivity. In order to obtain information about
the mean flux of the radio bright and radio faint samples, we combine
the data from many observations to form stacked observations using
visibility stacking \citep{hancock_visibility_2011}. In this and
subsequent sections we use a subset of the data listed in
\citet{chandra_radio-selected_2012}. The selection criteria for this
subset are discussed in the next section. Our analysis of
the stacking results is done twice; first, by simply appealing to the
large difference in flux between the two populations, and second, by
comparing the measured fluxes to predictions generated from our model
luminosity distributions. The two analyses come to the same conclusion
and a time-poor reader may skip section \ref{sec:modeling}.

\subsection{Visibility stacking}\label{sec:visibility_stacking}
Image based stacking has been used previously in astronomy to
investigate the mean properties of a population of objects which
cannot be easily detected individually. Traditionally, stacking
involves creating a calibrated image of each source under consideration
and then forming a weighted sum of these images. Under the assumption
of Gaussian noise that is uncorrelated between images and pixels, the
stacking of $N$ images will result in a factor of $\sim\sqrt{N}$
improvement in sensitivity. \citet{white_signals_2007} used image-based stacking to measure the mean radio flux of SDSS quasars in the
FIRST survey. They note that the interferometric nature of radio
images and the need for deconvolution produce spatially correlated
noise that makes it difficult to reach the ideal sensitivity of the
stacked images even when care has been taken to ensure a consistent
$(u,v)$ coverage. In particular, the mean of noisy data does not
converge to the true mean, and the relation between the stacked value
and the mean of the population depends on the structure of the
underlying noise in a non-linear manner. In
\citet{hancock_visibility_2011} we detailed the method of visibility
stacking, in which the calibrated visibility data is combined {\em
before} imaging takes place. Visibility stacking makes it possible to
stack radio observations with different $(u,v)$ coverages, and thus to
avoid problems associated with the structure of the underlying noise.

\label{sec:reduction}
In order to obtain an homogeneous sample, we selected the $8.46$\,GHz
observations from \citet{chandra_radio-selected_2012} from the VLA as
they comprise the largest subset of observations. The data were
obtained from the VLA archive, flagged and calibrated in \aips\,
\citep{greisen_AIPS_2003}, with a ParselTongue
\citep{kettenis_parseltongue:_2006} script based on that of
\citet{bell_automated_2011}, and then exported to \miriad\,
\citep{sault_retrospective_1995} for stacking and imaging.

Observations after 2006 routinely included one or more antennas with
expanded VLA \citep[JVLA,][]{perley_expanded_2011} receivers. All baselines including JVLA receivers
were flagged and not used in this analysis. Of the 999 observations
retrieved from the VLA archive, 226 were excluded due to calibration
problems that could not be resolved. The remaining 773 observations
were calibrated, imaged, and manually inspected for background sources
such as an active nucleus or HII regions within the host galaxy, or
other radio sources within the field of view. Excluding the GRB
afterglows, all radio sources were modeled and removed from the
visibility data, so that they would not contribute flux to the final
stacked observation. Observations that included complex sources that
were not able to be subtracted accurately were excluded from the
analysis. For 36 observations the background emission was not able to
be completely subtracted, and these observations were excluded from
our analysis. 

In total, 737 observations of 178 GRBs were used in this
work. The effective total integration time is $17.8$\,days, with
$13.2$\,days of observing time dedicated to GRBs that were detected in
at least one epoch, and only $4.6$\,days of observing time dedicated
to GRBs that were never detected. The difference in observing time
between the two samples reflects a typical observing strategy in which
a GRB is no longer observed after the first week if no detection has
been made, but is otherwise monitored regularly.

Observations that were suitable for stacking were binned into groups
depending on the time elapsed since the burst, and a stacked image was
created for each bin. {\bf The bin sizes were chosen to be a compromise
between: large bins that result in sensitive stacked images, and small bins in
which the radio afterglow does not evolve significantly. The time bins were
spaced logarithmically between $0.1$\,days and $200$\,days}. A
GRB is considered to be {\em bright} if at least one
observation resulted in a detection, and {\em faint} no detection was
ever made. Separate stacked observations were created for the radio
bright and radio faint GRBs. {\bf If a GRB is detected in at least one
observation, then all observations of this GRB will be included in the radio
bright stacked observation, even if a particular observation didn't result in
a detection}. Each of the stacked observations was then imaged.

A detection in a visibility stacked image will not resemble the point
spread function calculated from the visibility sampling function, even
if all the individual sources are unresolved. Instead, the shape of
the detection is a sum of the point spread functions of each
individual observation, weighted by the flux of each source
observed. As the flux of the individual sources is inherently unknown
it is not possible to reconstruct the expected ``dirty beam'' and thus
it is not possible to deconvolve the stacked image. We used the
BLOBCAT package \citep{hales_blobcat:_2012} to extract a meaningful
flux from the stacked observations in which a detection was made. The
sensitivity of each of the stacked observations was measured from the
pixel rms in the images. The sensitivity of the stacked images was
found to be worse than the theoretical sensitivity expected from a
single observation of equivalent integration time. We attribute this
non-ideal sensitivity improvement to the presence of faint background
sources within individual images, which were not able to be identified
or removed, as well as calibration errors (which are difficult to
detect in empty images). The stacked observations were more sensitive
than any of the individual observations, resulting in upper limits on
the mean flux of the population that were $4-8$ times fainter than any
of the individual observations.

\subsection{Preliminary analysis}\label{sec:prelim_analysis}
\begin{figure}
\centering
\includegraphics[width=0.95\linewidth]{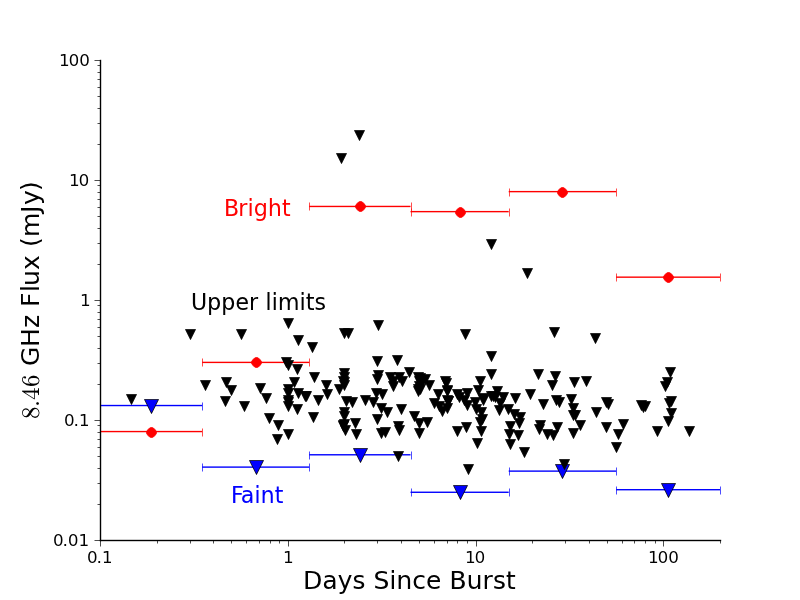}
\caption{The flux of the stacked observations for the radio bright
(red) and radio faint (blue, upper limits) GRBs. The black triangles
are the $3\sigma$ upper limits for the individual observations that
were used to create the stacked observation of the radio faint
GRBs. There is a factor of $10-1000$ difference between the stacked
flux of the bright GRBs and the stacked upper limit of the faint GRBs,
in all but the first time bin.}
\label{fig:stacked_overlay}
\end{figure}

Figure\,\ref{fig:stacked_overlay} shows the results of the visibility
stacking. The stacked observations of the radio bright GRBs resulted,
as expected, in strong detections at each epoch {\bf that are consistent with
the evolution of a canonical GRB afterglow}. However, the new result we present
here is that the stacked data of radio faint GRBs did not result in any
detections. The lack of a detection is inconsistent with the idea that the radio
faint GRBs are simply a fainter tail of the radio bright GRB population, subject
to limited observing sensitivity. The mean flux of the radio bright and radio
faint GRB populations differ by up to three orders of magnitude. Such a large
difference in flux suggests that the radio bright and radio faint GRBs are
intrinsically different. To confirm that this result is statistically
significant, we model the expected flux of the two
populations in Section \ref{sec:modeling}. 

\subsection{Population Modeling}\label{sec:modeling}
The faintest individual GRB detections and the typical upper-limits of
non-detected individual GRBs both occur at about the same flux (see
Figure\,\ref{fig:detectionrate}). This is consistent with the previous
interpretation that there is a single population of GRBs and that we
are currently only able to detect the brightest 30\% due to the
limited sensitivity of our telescopes. Our null hypothesis is thus
that {\bf all GRBs have a radio afterglow that, when taken together,}
form a single broad distribution in radio flux, and that the
difference between the radio bright and faint samples is an artifact
of observing sensitivity. In order to test this hypothesis we create
GRB afterglow models that are consistent with current observations and
use these models to predict the flux of the undetected GRB
afterglows. We then test these predictions using more sensitive
observations obtained from visibility stacking (see section
\ref{sec:visibility_stacking}).

{\bf Since the true distribution of GRB radio luminosities is unknown, we will create
three single peaked models for this distribution and then look at the range of radio
fluxes that these models predict}. Following \citet{berger_radio_2003}, we
describe the number of GRBs with $\log(L_{radio}[W/Hz])$ between $\ell$ and
$\ell+d\ell$ using three different, two parameter models as follows:

A distribution that is Gaussian in $\ell$, with mean $\ell_0$ and
variance $\sigma_\ell^2$:
\begin{equation}
n(\ell) = \frac{1}{\sqrt{2\pi\sigma_\ell}}\exp\left[-\frac{1}{2}\left(\frac{\ell-\ell_0}{\sigma_\ell} \right)^2\right],
\end{equation}
a flat distribution with luminosities between $\ell_1$ and $\ell_2$:
\begin{equation}
n(\ell) = \begin{cases}
0 & \mathrm{if}\, \ell<\ell_1\\
\frac{1}{\ell_2-\ell1} & \mathrm{if}\, \ell_1\leq \ell \leq \ell_2,\\
0 & \mathrm{if}\, \ell>\ell_2\\
\end{cases}
\end{equation}
and a decreasing power-law (DPL) with a lower cutoff of $\ell_0$ and exponent $\alpha_\ell$:
\begin{equation}
n(\ell)= \begin{cases}
0 & \mathrm{if}\, \ell<\ell_0\\
(1-\alpha_\ell)\ell^{\alpha_\ell}/\ell_0^{(\alpha_\ell+1)} & \mathrm{if}\, \ell\geq \ell_0\
\end{cases}.
\end{equation}

We convert the above luminosity distributions into a distribution of
observed fluxes using a model redshift distribution.  The distributions
of redshifts for GRBs that do or do not have radio detections are the
same, and the radio/optical/X-ray/gamma-ray properties of GRBs that do
and do not have a measured redshift are no different (see
\S\,\ref{sec:redshifts}). We therefore take the redshift distribution
of the combined (bright and faint) VLA-observed GRB sample as our
model distribution. The expected flux distribution can then be
calculated by combining the luminosity and redshift distributions such
that the number of GRBs with fluxes between $s$ and $s + ds$ is given
by:

\begin{equation}
n(s) = \mathcal{F}(s:n(\ell),n(z)),
\end{equation}
where the function $\mathcal{F}(\cdot)$ measures the expected number
of GRBs with fluxes between $s$ and $s+ds$, given a distribution of
$\ell=\log(L_{radio}[W/Hz])$, $n(\ell)$, and a distribution of redshifts, $n(z)$. We use
a cosmology parametrized by $H_0 =71$km/s/Mpc, $\Omega_m=0.27$, and
$\Omega_{vac}=0.73$.

We measure the goodness of fit for a given model $n(\ell)$ by computing the
likelihood function $\mathcal{L}$:

\begin{equation}
\mathcal{L}_j = \begin{cases}
\int_0^\infty n(s)G(s_j,\sigma_j)ds & \text{ for detections}\\
\int_0^\infty n(s)H(3\cdot\sigma_j)ds & \text{ for non-detections}\\
\end{cases}
\end{equation}
and
\begin{equation}
\mathcal{L}=\prod_j\mathcal{L}_j
\end{equation}

where $G(s_j,\sigma_j)$ is a normalized Gaussian centered on the
measured flux $s_j$ with a FWHM equal to the measurement uncertainty
$\sigma_j$, and $H(3\cdot\sigma_j)$ is a normalized step function that
is non-zero below, and zero above, the $3\sigma$ detection limit of the
observation. The index $j$ iterates over all the observations within
the given time bin.

We bin the observations into the same four time bins that were
described in section\,\ref{sec:visibility_stacking}. By maximizing
$\mathcal{L}$ for each of the luminosity distributions, we obtained
the most likely parameters for each of the luminosity models, for each
of the time bins. Shown in Table\,\ref{tab:model_params} are the
parameters for each model, $1.3-4.5$\,days post burst.

\begin{table}
\centering
\caption{Model parameters obtained from maximization of $\mathcal{L}$, for
observations $1.3-4.5$\,days post burst.}
\label{tab:model_params}
\begin{tabular}{lll}
n($\ell$) model & Parameter 1  & Parameter 2 \\
\hline
Gaussian  & $\ell_0$ = 19.6  & $\sigma_\ell$ = 0.6 \\
Flat      & $\ell_1$ = 18.6  & $\ell_2$ = 21.2 \\
DPL       & $\ell_0$ = 19.0  & $\alpha_\ell$ = -28.5 \\
\hline
\end{tabular}
\end{table}

\begin{figure}
\centering
\includegraphics[width=\linewidth]{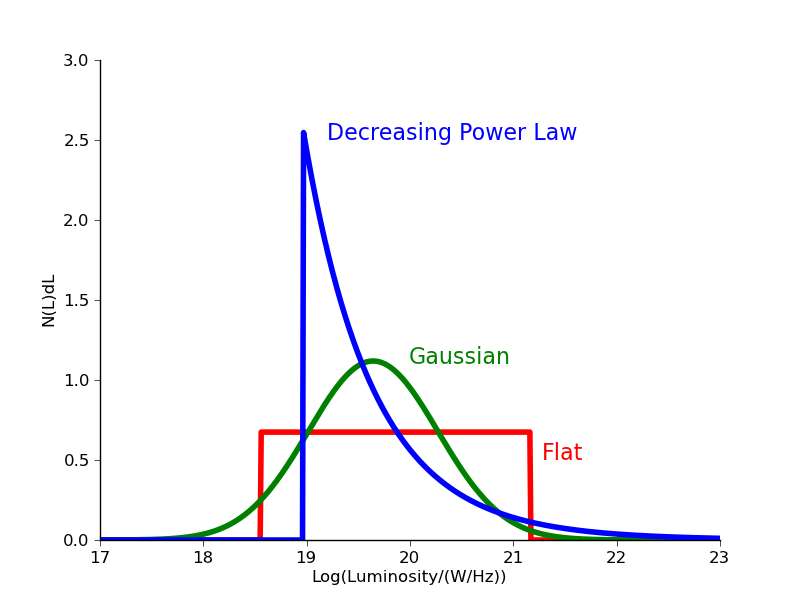}
\includegraphics[width=\linewidth]{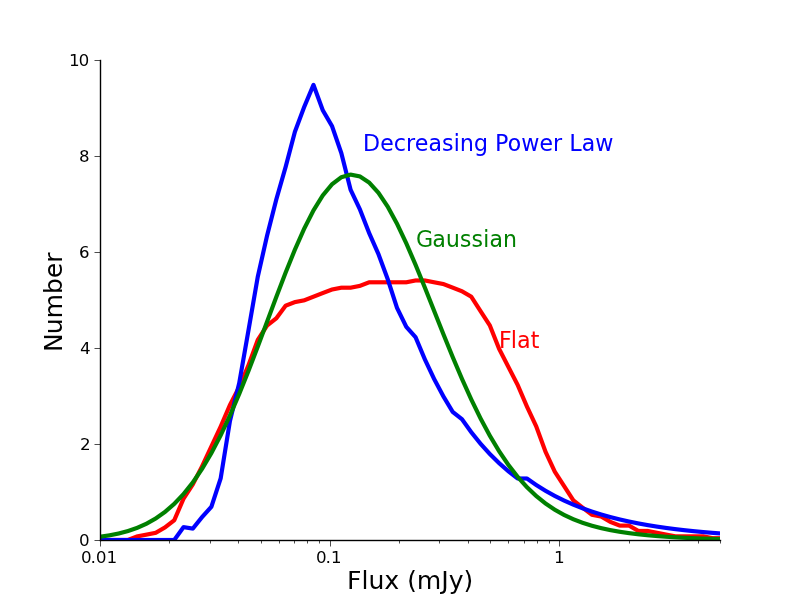}
\caption{The three model flux distributions with best fit parameters
as given in Table\,\ref{tab:model_params} for $1.3-4.5$\,days post
burst. {\em Top}: The luminosity distributions, {\em Bottom}: The
corresponding flux distributions when redshift has been taken into
account.}
\label{fig:models}
\end{figure}

\subsubsection{Model Predictions}\label{sec:predictions}
In Figure\,\ref{fig:models} we show the models for observations (1.3-4.5)\,days
post burst that maximize $\mathcal{L}$. By drawing fluxes from
the model distribution $n(s)$ and randomly assigning an observing
sensitivity drawn from the set of radio observations, we are able to
divide a model population of GRBs into radio bright (detected) and
radio faint (not detected) subsets. When averaged over repeated
drawings, these two subsets can then be used to determine the expected
detection rate, and the amount of flux we could expect to see in a
stacked observation. The three models predict detection rates of
between $20-50\%$ with uncertainties that are consistent with the
observed $30\%$.  All the models predict that the radio bright GRBs
should have a stacked flux of between $0.4$ and $15$mJy at
$8.46$\,GHz, depending on the model and time bin, whilst the radio
faint GRBs should have a stacked flux in the range
$0.09-0.14$\mJy. {\bf The first time bin ($0.1-0.35$\,days) contains only 7
observations, 3 of faint GRBs, and 4 of bright GRBs. Such a small number of
observations means that it is difficult to make accurate models or accurate
predictions of the expected fluxes. We therefore do not consider the first time
bin in our analysis}.

\begin{figure}
\centering
\includegraphics[width=\linewidth]{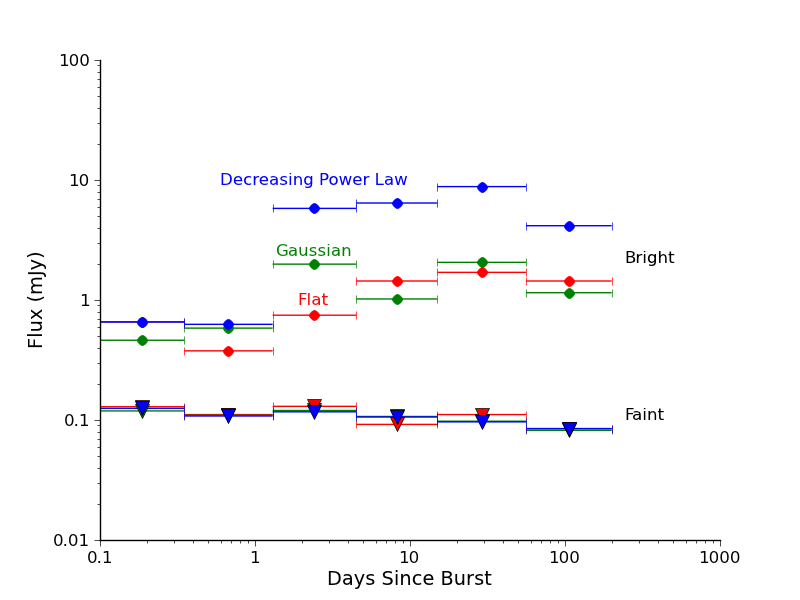}
\caption{The stacked flux of radio bright {\bf(upper circles)} and radio
faint {\bf(lower triangles)} GRB afterglows as predicted by each of the
three luminosity models.}
\label{fig:model_predictions}
\end{figure}

\begin{figure}
\centering
\includegraphics[width=\linewidth]{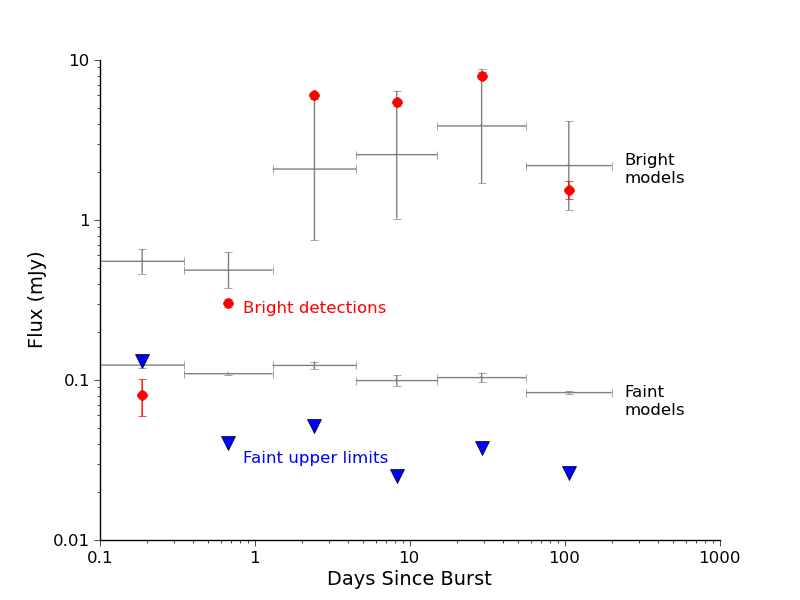}
\caption{The predicted stacked flux of the three models (in gray) are
over-plotted with the stacked flux of the bright GRBs (in red) and the
$3\sigma$ stacked upper limits on the faint GRBs (in blue). The vertical error
bars on the gray data points represent the range of fluxes predicted
by the models (cf. Fig\,\ref{fig:model_predictions}).  The models are
able to account for the radio bright population but substantially
over-predict the stacked flux of the radio faint population.}
\label{fig:model_comparison}
\end{figure}

Figure\,\ref{fig:model_predictions} shows the
predicted fluxes from each of the models for both the bright and faint
GRBs. In Figure\,\ref{fig:model_comparison} the range of predicted
fluxes are compared with the stacked observations. The radio bright
stacked observations result in a mean flux that is consistent with the
predicted range. The radio faint stacked observations result in an
upper limit that is five times fainter than the predicted range. The
fact that none of the flux models are able to account for the radio
faint GRBs is inconsistent with the hypothesis that there is a single
broad distribution of GRB fluxes that result in a sensitivity limited
detection rate.

We have now shown that neither redshift nor observational sensitivity
are responsible for the low detection rate of GRB radio
afterglows. The division of GRBs into radio faint and radio bright is
therefore physical and must be due to intrinsic differences
between the two populations.

\subsection{Refined analysis}\label{sec:contamination}
{\bf
The null hypothesis stated that all GRBs produce a radio bright afterglow which
taken together, form a single peaked distribution of fluxes. We have show that
this hypothesis is not supported by the data, and therefore that GRBs without a
detected afterglow must, at least in part, be truly radio faint. There is of
course some amount of contamination in what we call the radio faint sample,
which could be overcome with better observing sensitivity. To understand the true
fraction of GRBs that are radio bright and faint, we calculate the fraction of
radio bright GRBs that are within our radio faint sample. For GRBs observed
$0.35-1.3$\,days post burst, our models predict a mean flux of $0.19$\,mJy,
whereas the stacked upper limit is $0.04$\,mJy. Therefore it is possible for
$21$\% of the faint GRB sample to have fluxes drawn from the radio bright
distribution, and still be consistent with the stacked upper limit. Since $59$\%
of the GRBs observed in this time bin are in the radio faint population the true
(total) fraction of radio bright GRBs is $(59\%+21\%\times41\%=) \sim 70\%$.

The above analysis assumes that observed GRBs are a representative sample, which will
be the case during the first week or two after the burst. At later times, GRBs
with an established afterglow will be monitored, where as those without an
afterglow are likely to be ignored. Table\,\ref{tab:true_detection} shows the
fraction of radio bright and radio faint GRBs observed in each time bin
(\%observed) as well as the calculated true fraction of radio bright GRBs
(\%corrected). The late time observing bias can be seen in the increasing
fraction of radio bright GRBs observed. However with the exception of the final
time bin, the corrected fraction of true radio bright GRBs remains between
$60-70\%$. We therefore conclude that the true fraction of radio faint GRBs is
only $30-40\%$.}

\begin{table}
\centering
\caption{The fraction of radio bright GRBs in each of the time bins, either as
observed, or when corrected for contamination. See text for details.}
\label{tab:true_detection}
\begin{tabular}{ccc}
\hline
days        & \multicolumn{2}{c}{radio bright GRBs}\\
since burst & \%observed & \%corrected \\
\hline
0.35-1.3 & 41\% & 67\% \\
0.3-4.5  & 47\% & 73\% \\
4.5-15   & 52\% & 60\% \\
15-56    & 65\% & 60\% \\
56-200   & 81\% & 44\% \\
\hline
\end{tabular}
\end{table}

\section{Multi-wavelength properties of the two GRB populations}\label{sec:multi-wavelength}
We now turn to the multi-wavelength properties of our sample of GRBs to
investigate the possible cause of the two populations. {\bf The sample of GRBs that
we have considered are a subset of the \citet{chandra_radio-selected_2012} GRBs.
\citet{chandra_radio-selected_2012} found a consistent and significant
difference between the multi-wavelength properties of the radio bright and radio
faint GRBs. To verify that our selection criteria has produced a
representative sample of the complete data, we perform the same analysis on our
subset of the data.}

\begin{figure}
\centering
\includegraphics[width=0.95\linewidth]{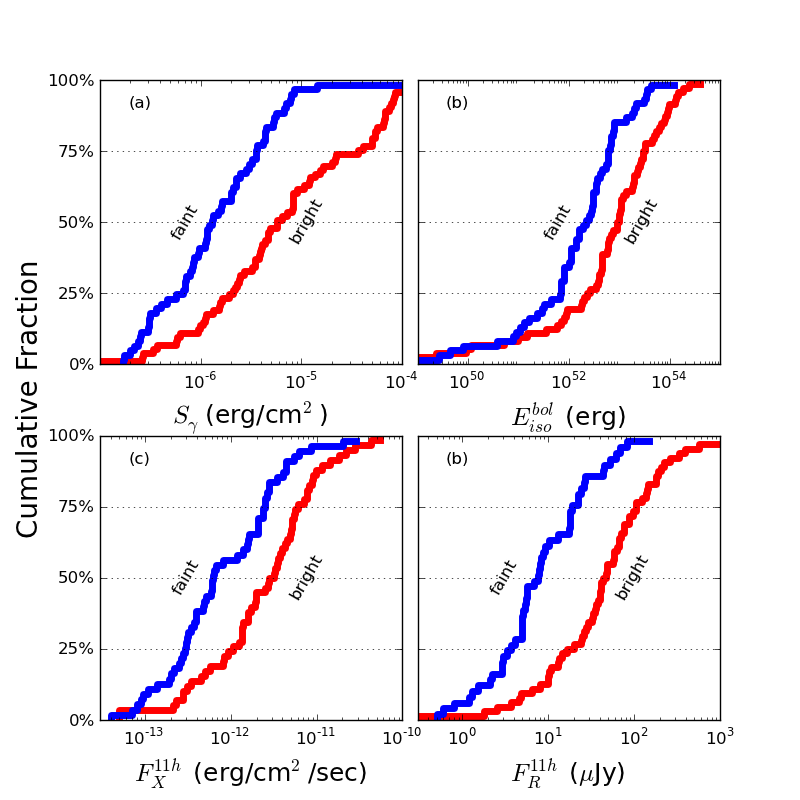}
\caption{Cumulative distribution functions for GRBs with radio
bright vs radio faint afterglows. The properties shown are: a - gamma
ray fluence, b - isotropic energy release, c - X-ray flux at 11h, and
d - optical flux at 11h. Each of these parameters show a significant
difference between the two radio populations as reported in Table
\ref{tab:medians}.}
\label{fig:cdf}
\end{figure}

In Figure\,\ref{fig:cdf} we show the distribution of four different measures of
brightness from optical to gamma-rays. Table\,\ref{tab:medians} presents the
median values of these properties as well as the redshift and \tninety\, for the
radio bright and radio faint populations. {\bf As many as $20-40\%$ of the GRBs
in our radio faint sample may actually be radio bright (see
\S\,\ref{sec:contamination}) and yet we are still able to detect a significant
difference between the radio bright and radio faint samples.} The radio faint
GRBs are consistently fainter than the radio bright GRBs in each of the measures
of brightness at other wavelengths, consistent with the findings of
\citep{chandra_radio-selected_2012}.

\begin{figure}
\centering
\includegraphics[width=0.95\linewidth]{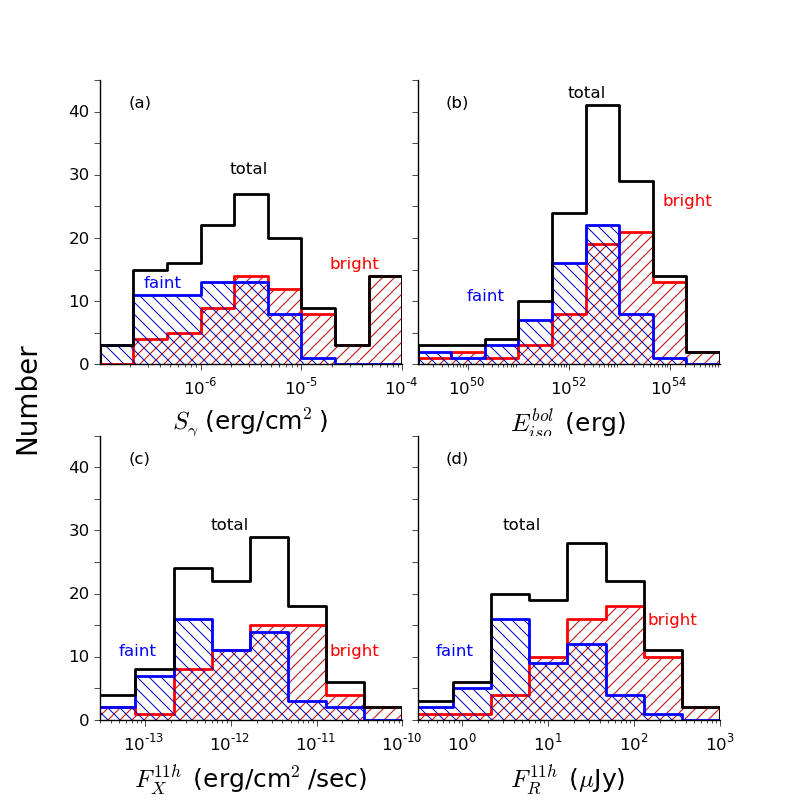}
\caption{Histograms comparing the properties of GRBs with radio bright
and radio faint afterglows. The combined population is shown in
black. The properties shown are the same as in
Figure\,\ref{fig:cdf}. The difference between the populations is
significant, but the magnitude of the difference is not evident
when shown as a histogram.}
\label{fig:pdf}
\end{figure}

The difference between the two populations is both significant, and consistent. 
However, at wavelengths shorter than the radio, the difference is only a
factor of a few. {\bf In Figure\,\ref{fig:pdf} we plot a (more traditional) histogram of
the data in Figure\,\ref{fig:cdf}}. Due to the small number of known GRBs
and the large spread in their brightness, the histograms necessarily have bin
sizes that are similar to the difference between the two populations. It is this
combination of GRB number, spread in brightness and choice of plotting technique
that could otherwise lead one to overlook the difference between the two
populations.

\begin{table}
\centering
\caption{The median properties of the radio bright and radio faint
GRBs. The final column is the K-S statistic p-value.}
\label{tab:medians}
  \begin{tabular}{lrrc}
    \hline
&\multicolumn{2}{c}{Population}\\
Parameter (median)                    & bright   & faint    & p     \\
\hline
redshift                              & 1.4      & 1.3      &0.32   \\
\tninety\, (s)                        & 62       & 34       &8.3e-3 \\
\fluence\, ($\times10^{-6}$erg/cm$^2$)& 5.7      & 1.6      &1.5e-5 \\
\fx\, ($\times10^{-13}$erg/cm$^2$/s)  & 23       & 6.4      &8.9e-5 \\
\fr\, ($\mu$Jy)                       & 41       & 5.8      &6.0e-11\\
\ebol\, ($\times10^{52}$erg)          & 10       & 2.1      &4.8e-5 \\
\hline
  \end{tabular}
\end{table}

\section{Interpretation of the two populations}\label{sec:interpretation}
Well studied samples of GRBs \citep[eg, the gold samples
of][]{Tsutsui_universal_2013, Zhang_discerning_2009} have been influential in
developing the prompt and afterglow theory of GRBs. However it has been
implicitly assumed that all GRBs are radio bright. It is thus not surprising
that the standard model of GRB afterglows does not accurately describe the
properties of the radio faint population. We have concluded that there must be
two populations of GRBs, with different explosion mechanisms, radiation
processes, or environments, which are responsible for the different radio
fluxes. {\bf Our modeling and analysis suggest that $30-40\%$ of GRBs are truly
radio faint and $60-70\%$ are radio bright.} In this section we present two
possible explanations for the underlying physical differences between these two
populations of GRBs.

\subsection{Gamma-ray efficiency}
The partitioning of the total energy released by the GRB central
engine into prompt emission \ebol, and afterglow emission \eke, can be
parametrized by \epsgamma{} (the gamma-ray efficiency) as:

\[
\epsgamma = \frac{\ebol}{\ebol+\eke}
\]

The measured values of \epsgamma{} are found to vary greatly from as
little as 0.03 \citep{berger_nonrelativistic_2004}, to 0.5
\citep{granot_implications_2006}, and even as high as 0.9
\citep{nousek_evidence_2006}. Such a large variation in \epsgamma{}
(and hence the ratio of \ebol{} to \eke{}) means that it is not
possible to use \ebol{} to predict \eke{} and thus of the strength of
the radio afterglow even for those that are radio bright. The number
of GRBs with a measured \epsgamma{} is large enough to show that there
are both large and small values of \epsgamma{}, however, there are not
yet enough measurements to distinguish between a bimodal and
quasi-uniform distribution. It is possible that the two populations of
radio bright and radio faint GRBs that we have identified in the
previous section are representative of GRBs with either low
\epsgamma{} (radio bright), or high \epsgamma{} (radio faint). The
difference in \epsgamma{} could be due to either differences in the
emission mechanism, or the nature of the central engine.

\subsubsection{Prompt emission mechanism}
The underlying bi-modality of \epsgamma{} could be a
result of different emission mechanisms that are predicted by the
different prompt emission models such as the the electromagnetic model
\citep[EMM,][]{lyutikov_electromagnetic_2006}, or the fireball model
\citep[FBM,][]{piran_gamma-ray_1999}. The EMM with a very low baryon
loading can generate intense prompt emission with a large \epsgamma{},
meaning that the afterglow will be faint or non-existent. The standard
FBM involves an intermediate baryon loading that will result in a low 
\epsgamma{} and a radio bright afterglow.

\subsubsection{Central engine}
Even within the FBM, it is possible to obtain two populations with low and high
values of \epsgamma{} which are in turn the root cause of the radio bright and
faint GRB afterglow populations, respectively. \citet{Komissarov_shock_2012} has
shown that the fraction of energy radiated in the prompt phase (effectively
\epsgamma{}) is inversely proportional to strength of the magnetic field
produced by the central engine. Stronger magnetic fields produce less efficient
prompt emission and thus $\epsgamma \propto 1/B$. {\bf Whilst black holes are
the favored candidate for most GRB central engines, millisecond magnetars have
been proposed as another possibility \citep{Zhang_open_2011}}. The magnetic
field strength of a milli-second magnetar ($\sim 10^{14-15}G$) would be much
greater than that at the innermost stable circular obit of a similar mass black
hole \citep[$\lesssim 10^8G$,][]{Piotrovich_magnetic_2010}. Thus two populations
of GRBs, one magnetar-driven, and one black-hole-driven, {\bf could perhaps
provide a natural explanation} for two populations of \epsgamma{} and could give
rise to the radio bright and radio faint GRB populations, respectively, which we
observe.

The claimed observational signature for a magnetar-driven central engine, is the
presence of an X-ray plateau that ends with a sharp decay. Ten long GRBs have
been identified by \citet{troja_swift_2007}, \citet{dallosso_gamma-ray_2011},
and \citet{obrien_observational_2011} {\bf for which this X-ray signature is
potentially present}. Of these 10 GRBs, only 5 were observed at radio
frequencies with two being detected (GRB\,061121A, GRB\,071021A) and three not
being detected. {\bf The small number of measurements prevents any definitive
conclusions. However, should the observed trend be maintained in further
observations, this would argue that magnetar-driven central engines are probably
not responsible for radio bright afterglows.}

\subsection{Observational outcomes}
If differences in \epsgamma{} lead to radio bright and radio faint GRB
afterglows, then the radio bright low-luminosity GRBs (llGRBs) are
interesting in that they hint at a population of fainter GRBs that are
below the detection limits of \swift\, but which have radio afterglows
detectable with our current generation of radio telescopes. Such a
population of gamma-ray faint GRBs would bridge the gap between llGRBs
and engine driven supernovae \citep{soderberg_relativistic_2010}. The
non-detection of such a population, as yet, is not surprising given
the current lack of wide-field, sensitive, transient radio
surveys. However, upcoming projects such as the variable and slow
transients survey \citep[VAST,][]{murphy_vast:_2013} which make use of
large field of view radio observations will be able to detect the
afterglow of such a population, and optical transient surveys such as
the panoramic survey telescope and rapid response system
\citep[Pan-STARRS,][]{kaiser_pan-starrs:_2002}, the Palomar Transients
Factory \citep[PTF,][]{law_palomar_2009}, SkyMapper
\citep{keller_skymapper_2007}, or the Antarctic Schmidt Telescopes
\citep{yuan_progress_2010} should be able to detect the prompt optical
signature of these objects.

{\bf Regardless of the cause of difference between the radio bright and radio
faint GRB populations, we predict that future GRB radio observations with an rms
of $\sim 10\,\mu$Jy will result in an a detection rate as high as $60-70\%$, but
not higher. This rms is typical of observations made with the JVLA
\citep[eg,][]{corsi_GRB_2013}. A preliminary analysis of GRB observations with
the JVLA as reported through the GCN circular archive \citep{barthelmy_GRB_2000}
reveals a detection rate of $60\%$ for $2012-2013$, which is in agreement with
this analysis.}

\section{Conclusions}
We have taken a sample of 737 observations of 178 GRBs from the VLA, and found
that the difference between the detected (radio bright) and non-detected (radio
faint) GRB radio afterglows is not simply a result of observing sensitivity. By
stacking the radio observations we find that the radio faint GRBs are not a low
luminosity tail of the radio bright population but a second population of GRBs
that are intrinsically less luminous at all wavelengths. We suggest that the
radio faint population is a result of high gamma-ray efficiency, resulting from
different prompt emission mechanisms or different central engines. These
possibilities will be explored in future work. {\bf Approximately 1 in every 3
GRBs are radio faint, and future theoretical work will need to
consider such a population.}

\section*{Acknowledgments}
We thank Davide Burlon, Jochen Greiner, Chryssa Kouveliotou, and Dale
Frail for helpful conversations and suggestions, and John Benson
for assistance with accessing large amounts of the VLA archive. {\bf We also
thank the anonymous referee for suggestions that significantly improved this
work.}

The National Radio Astronomy Observatory is a facility of the National
Science Foundation operated under cooperative agreement by Associated
Universities, Inc. This research has been supported by the Australian
Research Council through Super Science Fellowship grant
FS100100033. The Centre for All-sky Astrophysics is an Australian
Research Council Centre of Excellence, funded by grant CE110001020.

\label{lastpage}
\input{journal_short.def}
\bibliographystyle{apj}
\bibliography{ms}
\end{document}